\documentclass[traditabstract]{aa}  
\usepackage{graphicx}
\usepackage{txfonts}

\begin{document}

\title{Hot subdwarf stars in close-up view}

\subtitle{II. Rotational properties of single and wide binary subdwarf B stars 
\thanks{Based on observations at the Paranal Observatory of the European 
Southern Observatory for programmes number 165.H-0588(A), 167.D-0407(A), 071.D-0380(A) and 072.D-0487(A).
Based on observations at the La Silla Observatory of the 
European Southern Observatory for programmes number 073.D-0495(A), 074.B-0455(A), 076.D-0355(A), 077.D-0515(A) and 078.D-0098(A).
Based on observations collected at the Centro Astron\'omico Hispano Alem\'an (CAHA) at Calar Alto, operated jointly by the Max-Planck Institut f\"ur Astronomie and the Instituto de Astrof\'isica de Andaluc\'ia (CSIC). Some of the data presented here were obtained at the W.M. Keck Observatory, which is operated as a scientific 
 partnership among the California Institute of Technology, the University of California, and the National Aeronautics and Space Administration. The Observatory was made possible by the generous financial support of the W.M. Keck Foundation. Based on data obtained with the Hobby-Eberly Telescope (HET), which is a joint project of the University of Texas at Austin, the Pennsylvania State University, Stanford University, Ludwig-Maximilians-Universit\"at M\"unchen, and Georg-August-Universit\"at G\"ottingen.}
}

\author{S. Geier \inst{1}
   \and U. Heber \inst{1}
   }

\offprints{S.\,Geier,\\ \email{geier@sternwarte.uni-erlangen.de}}

\institute{Dr. Karl Remeis-Observatory \& ECAP, Astronomical Institute,
Friedrich-Alexander University Erlangen-Nuremberg, Sternwartstr. 7, D 96049 Bamberg, Germany }

\date{Received \ Accepted}

\abstract{Subluminous B stars (sdBs) form the extremely hot end of the horizontal branch and are therefore related to the blue horizontal branch (BHB) stars. While the rotational properties of BHB stars have been investigated extensively, studies of sdB stars have concentrated on close binaries that are influenced by tidal interactions between their components. Here we present a study of 105 sdB stars, which are either single stars or in wide binaries where tidal effects become negligible. The projected rotational velocities have been determined by measuring the broadening of metal lines using high-resolution optical spectra. All stars in our sample are slow rotators (${v_{\rm rot}\sin{i}}<10\,{\rm km\,s^{-1}}$). Furthermore, the $v_{\rm rot}\sin{i}$-distributions of single sdBs are similar to those of hot subdwarfs in wide binaries with main-sequence companions as well as close binary systems with unseen companions and periods exceeding $\simeq1.2\,{\rm d}$. We show that blue horizontal and extreme horizontal branch stars are also related in terms of surface rotation and angular momentum. Hot blue horizontal branch stars ($T_{\rm eff}>11\,500\,{\rm K}$) with diffusion-dominated atmospheres are slow rotators like the hot subdwarf stars located on the extreme horizontal branch, which lost more envelope and therefore angular momentum in the red-giant phase. The uniform rotation distributions of single and wide binary sdBs pose a challenge to our understanding of hot subdwarf formation. Especially the high fraction of helium white dwarf mergers predicted by theory seems to be inconsistent with the results presented here.
}

\maketitle

\section{Introduction \label{sec:intro}}

Subluminous B stars (sdBs) show similar colours and spectral characteristics to main sequence stars of 
spectral type B, but are less luminous. Compared to main sequence B stars, the hydrogen Balmer lines in the spectra 
of sdBs are stronger while the helium lines are much weaker. The strong line broadening and the early confluence of the
Balmer series is caused by the high surface gravities ($\log\,g\simeq5.0-6.0$) of these compact stars 
($R_{\rm sdB}\simeq0.1-0.3\,R_{\rm \odot}$). Subluminous B stars are considered to be core helium-burning stars with 
very thin hydrogen envelopes and masses of about half a solar mass (Heber \cite{heber86}) located at the extreme end of the horizontal branch (EHB). 

\subsection{Hot subdwarf formation \label{sec:formation}}

The origin of EHB stars is still unknown (see Heber
\cite{heber09} for a review). The key question is how all but a tiny fraction of the red-giant progenitor's hydrogen envelope was removed at the same time at which the helium core has attained the mass ($\simeq0.5\,M_{\rm \odot}$) to ignite the helium flash. The reason for this high mass loss at the tip of the red giant branch (RGB) is unclear. Several single-star scenarios are under discussion (D'Cruz et al. \cite{dcruz96}; Sweigart \cite{sweigart97}; De Marchi \& Paresce \cite{demarchi96}; Marietta et al. \cite{marietta00}), which require either a fine-tuning of parameters or extreme environmental conditions that are unlikely to be met for the bulk of the observed subdwarfs in the field. 

According to Mengel et al. (\cite{mengel76}), the required strong mass loss can occur in a close-binary system. The progenitor of the sdB star has to fill its Roche lobe near the tip of the red-giant branch (RGB) to lose a large part of its hydrogen envelope. The merger of close  binary white dwarfs was investigated by Webbink (\cite{webbink84}) and Iben \& Tutukov (\cite{iben84}), who showed that an EHB star can form when two helium core white dwarfs (WDs) merge and the product is sufficiently massive to ignite helium. Politano et al. (\cite{politano08}) proposed that the merger of a red giant and a low-mass main-sequence star during a common envelope (CE) phase may lead to the formation of a rapidly rotating single hot subdwarf star.

Maxted et al. (\cite{maxted01}) determined a very high fraction of radial velocity variable sdB stars, indicating that about two thirds of the sdB stars in the field are in close binaries with periods of less than 30 days (see also Morales-Rueda et al. \cite{morales03}; Napiwotzki et al. \cite{napiwotzki04a}; Copperwheat et al. \cite{copperwheat11}). Han et al. (\cite{han02,han03}) used binary population synthesis models to study the stable Roche lobe overflow (RLOF) channel, the common envelope ejection channel, where the mass transfer to the companion is dynamically unstable, and the He-WD merger channel.

The companions are mostly main sequence stars or white dwarfs. If the white dwarf companion is sufficiently massive, the merger of the binary system might exceed the Chandrasekhar mass and explode as a type Ia supernova. Indeed, Maxted et al. (\cite{maxted00}) found the sdB+WD binary KPD\,1930$+$2752 to be a system that might qualify as a supernova Ia progenitor (see also Geier et al. \cite{geier07}). In Paper~I of this series (Geier et al. \cite{geier10b}) more candidate systems with massive compact companions, either massive white dwarfs or even neutron stars and black holes, have been found. Furthermore, Geier et al. (\cite{geier11c}) reported the discovery of an eclipsing sdB binary with a brown dwarf companion.

\begin{figure}[t!]
\begin{center}
	\resizebox{\hsize}{!}{\includegraphics{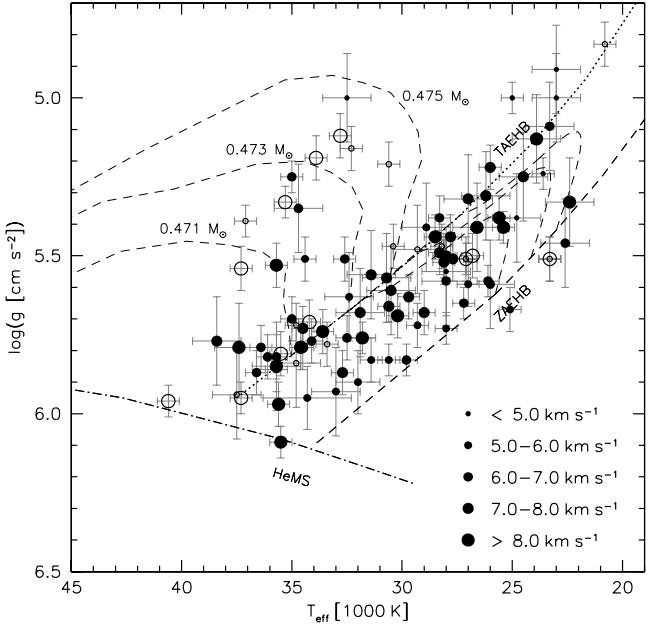}}
	\caption{$T_{\rm eff}-\log{g}$-diagram for the entire sample (not RV-variable) under study. 
	 The helium main sequence (HeMS) and the EHB band (limited by the zero-age 
         EHB, ZAEHB, and the terminal-age EHB, TAEHB) are superimposed with EHB evolutionary tracks for solar metallicity taken from 
	Dorman et al. (\cite{dorman93}) labelled with their masses. Open circles mark objects where only upper limits could be derived for $v_{\rm rot}\sin{i}$, filled circles objects with significant $v_{\rm rot}\sin{i}$. The size of the symbols scales with the value of $v_{\rm rot}\sin{i}$.}
	\label{fig:tefflogg}
\end{center}
\end{figure}

\subsection{Rotation on the horizontal branch \label{sec:rotation}}

The rotational properties of horizontal branch (HB) stars both in globular clusters and in the field all the way from the red to the blue end  have been studied extensively in the last decades (Peterson \cite{peterson83b}, \cite{peterson85}; Peterson et al. \cite{peterson83a}, \cite{peterson95}; Behr et al. \cite{behr00a}, \cite{behr00b}; Kinman et al. \cite{kinman00}; Recio-Blanco et al. \cite{recio02}, \cite{recio04}; Behr \cite{behr03a}, \cite{behr03b}; Carney et al. \cite{carney03}, \cite{carney08}). Most of these investigations were motivated by the puzzling horizontal branch morphologies in some globular clusters and the search for second or third parameters responsible for this phenomenon. The most interesting result of these studies is the discovery of a significant change in the rotational velocities of blue horizontal branch (BHB) stars when their effective temperatures exceed $\simeq11\,500\,{\rm K}$. HB stars cooler than this threshold value show ${v_{\rm rot}\sin\,i}$ values up to $40\,{\rm km\,s^{-1}}$, while the hotter stars rotate with velocities lower than $\simeq10\,{\rm km\,s^{-1}}$. 

The transition in rotational velocity is accompanied by a jump towards brighter magnitudes in the colour-magnitude diagram (Grundahl et al. \cite{grundahl99}) and a change in the atmospheric abundance pattern. Stars cooler than $\simeq11\,500\,{\rm K}$ show the typi\-cal abundances of their parent population (e.g. For \& Sneden \cite{for10}), while stars hotter than that are in general depleted in helium and strongly enriched in iron and other heavy elements such as titanium or chromium. Lighter elements such as magnesium and silicon on the other hand have normal abundances (Behr et al. \cite{behr03a,behr03b}; Fabbian et al. \cite{fabbian05}; Pace et al. \cite{pace06}). Diffusion processes in the stellar atmosphere are most likely responsible for this effect. Michaud et al. (\cite{michaud83}) predicted such abundance patterns before the anomalies were observed (see also Michaud et al. \cite{michaud08}). Caloi (\cite{caloi99}) explained the sharp transition between the two abundance patterns as the disappearance of subsurface convection layers at a critical temperature. Sweigart (\cite{sweigart02}) indeed found that thin convective layers below the surface driven by hydrogen ionization should exist and shift closer to the surface when the effective temperature increases. At about $12\,000\,{\rm K}$ the convection zone reaches the surface and the outer layer of the star becomes fully radiative. Since convection is very efficient in mixing the envelope, diffusion processes do not operate in HB star atmospheres of less than $12\,000\,{\rm K}$. 

Slow rotation is considered as a prerequisite for diffusion. Michaud (\cite{michaud83}) was the first to show that meridional circulation stops the diffusion process as soon as the rotational velocity reaches a critical value and could explain the chemical peculiarity of HgMn stars in this way. Quievy et al. (\cite{quievy09}) performed similar calculations for BHB stars and showed that the critical rotational velocity is somewhere near $\simeq20\,{\rm km\,s^{-1}}$ at the transition temperature of $11\,500\,{\rm K}$. This means that the atmospheric abundances of stars with lower ${v_{\rm rot}\sin\,i}$ should be affected by diffusion processes.

What causes the slow rotation that allows diffusion to happen, is still unclear. Sills \& Pinsonneault (\cite{sills00}) used a standard stellar evolution code and modelled the distribution of rotational velocities on the BHB. In order to reproduce the two populations of fast and slow rotators they assumed two distinct main sequence progenitor populations with different rotational veloci\-ties. In their picture the slowly rotating BHBs originate from slowly rotating main sequence stars. 

Another possible explanation is the spin-down of the surface layers by diffusion itself. Sweigart (\cite{sweigart02}) argued that the radiative levitation of iron triggers a weak stellar wind that carries away angular momentum. Vink \& Cassisi (\cite{vink02}) showed that such winds are radiatively driven. 

Brown (\cite{brown07}) used a stellar evolution code including rotation and modelled the distribution of rotational velocities on the BHB. This code allows one to follow the evolution of the progenitor star through the He-flash. Brown (\cite{brown07}) argues that no signifi\-cant angular momentum is exchanged between the stellar core and stellar envelope during the flash. The surface rotation of their models highly depends on the rotation of the surface convection zone, which contains most of the outer envelope's angular momentum. Hot BHB stars without surface convection zone rotate slower than the cooler ones with convection zone. This approach allows one to reproduce the observed ${v_{\rm rot}\sin\,i}$-distribution of BHB stars without assuming bimodal stellar po\-pulations (Brown et al. \cite{brown08}).

While the rotational properties of horizontal branch stars both in 
globular clusters and in the field are thoroughly examined, the investigation of EHB stars has mostly been restricted to close binary systems, where tidal interaction plays a major role (Geier et al. \cite{geier10b}). Very few apparently single EHB stars have been studied so far, all of which are slow rotators ($<10\,{\rm km\,s^{-1}}$, e.g. Heber et al. \cite{heber00};  Edelmann \cite{edelmann01}).

In this paper we determine the projected rotational velocities of more than a hundred sdB stars by measuring the broadening of metal lines. In Paper~I (Geier et al. \cite{geier10b}) the rotational properties of sdBs in close binary system were derived and used to clarify the nature of their unseen companions. Here we focus on the rotational properties of apparently single sdBs and wide binary systems, for which tidal interactions become negligible. 

In Sect.~\ref{sec:obs} we give an overview of the observations of high-resolution spectra and the atmospheric parameters of our sample. The determination of the rotational properties of 105 sdB stars are described in Sect.~\ref{sec:rotlow}, the results are interpreted in Sect.~\ref{sec:distrib} and compared to the corresponding results for BHB stars in Sect.~\ref{sec:bhb}. The implications for the sdB formation scenarios and the further evolution to the white dwarf cooling tracks are discussed in Sect.~\ref{sec:implications} and Sect.~\ref{sec:wd}, respectively. Finally, a summary is given in Sect.~\ref{sec:summary}.

\section{Observations and atmospheric parameters \label{sec:obs}}

ESO-VLT/UVES spectra were obtained in the course of the ESO Supernovae Ia 
Progenitor Survey (SPY, Napiwotzki et al. \cite{napiwotzki01, napiwotzki03}) 
at spectral resolution $R\simeq20\,000-40\,000$ covering 
$3200-6650\,{\rm \AA}$ with two small gaps at $4580\,{\rm \AA}$ and 
$5640\,{\rm \AA}$. Each of the 50 stars was observed at least twice (Lisker et al. \cite{lisker05}). 

Another sample of 46 known bright subdwarfs was observed with the 
FEROS spectrograph ($R=48\,000$, $3750-9200\,{\rm \AA}$) mounted at the ESO/MPG 
2.2m telescope (Geier et al. \cite{geier12}). 

Six stars were observed with the FOCES spectrograph 
($R=30\,000$, $3800-7000\,{\rm \AA}$) mounted at the CAHA 2.2m telescope (Geier et al. \cite{geier12}).
 Two stars were observed with the HIRES instrument ($R=45\,000$, 
 $3600-5120\,{\rm \AA}$) mounted at the Keck telescope (Heber et al. \cite{heber00}). 
 One star was observed with the HRS fiber spectrograph at the Hobby Eberly Telescope ($R=30\,000$, $4260-6290\,{\rm \AA}$, Geier et al. \cite{geier10b}).

Because a wide slit was used in the SPY survey and the seeing
disk did not always fill the slit, the instrumental profile of some of the UVES spectra was seeing-dependent. 
This has to be accounted for to estimate the instrumental resolution (see Paper~I). 
The resolution of the spectra taken with the fiber spectrographs FEROS and FOCES was assumed to be constant. 

The single spectra of all programme stars were radial-velocity (RV) corrected and co-added in 
order to achieve higher signal-to-noise.

Atmospheric parameters of the stars observed with UVES have been determined by Lisker et al. (\cite{lisker05}). HD\,205805 and Feige\,49 have been analysed by Przybilla et al. (\cite{przybilla06}), the two sdB pulsators KPD\,2109+4401 and PG\,1219+534 by Heber et al. (\cite{heber00}), and the sdB binaries PG\,1725+252 and TON\,S\,135 by Maxted et al. (\cite{maxted01}) and Heber (\cite{heber86}), respectively. The rest of the sample was analysed in Geier et al. (\cite{geier12}) and a more detailed publication of these results is in preparation. We adopted the atmospheric parameters given in Saffer et al. (\cite{saffer94}) for $[$CW83$]$\,1758$+$36. 

The whole sample under study is listed in Tables~\ref{tab:vrot} and \ref{tab:vrotrv} and the effective temperatures are plotted versus the surface gravities in Fig.~\ref{fig:tefflogg}. Comparing the positions of our sample stars to evolutionary tracks, we conclude that all stars are concentrated on or above the EHB, which is fully consistent with the theory. We point out that the inaccuracies in the atmospheric parameters do not significantly affect the derived projected rotational velocities.

\section{Projected rotational velocities from metal lines 
\label{sec:rotlow}}

To derive $v_{\rm rot}\,\sin{i}$, we compared the observed spectra 
with rotationally broadened, synthetic line profiles using a semi-automatic 
analysis pipeline. The profiles were computed for the appropriate atmospheric parameters using the LINFOR program (developed by Holweger, Steffen and Steenbock at Kiel university, modified by Lemke \cite{lemke97}).

For a standard set of up to 187 unblended metal lines from 24 different ions and with
 wavelengths ranging from $3700$ to $6000\,{\rm \AA}$, a model grid with 
 appropriate atmospheric parameters and different elemental abundances was 
 automatically generated with LINFOR. The actual number of lines used as input 
 for an individual star depends on the wavelength coverage. Owing to the 
 insufficient quality of the spectra and the pollution with telluric features 
 in the  regions blueward of $3700\,{\rm \AA}$ and redward of 
 $6000\,{\rm \AA}$ we excluded them from our analysis. A simultaneous fit of 
 elemental abundance, projected rotational velocity and radial velocity was
  then performed separately for each identified line using the FITSB2 
  routine (Napiwotzki et al. \cite{napiwotzki04b}). A detailed investigation of statistical and systematic 
  uncertainties of the techniques applied is presented in Paper~I. Depending on the quality of the data and
  the number of metal lines used, an accuracy of about $1.0\,{\rm km\,s^{-1}}$ can be achieved. 
 For the best spectra with highest resolution the detection limit is about $5.0\,{\rm km\,s^{-1}}$. 

Projected rotational velocities of 105 sdBs have been measured (see Tables~\ref{tab:vrot}, \ref{tab:vrotrv}). Ninety-eight sdBs do not show any RV variability. In addition, seven are radial velocity variable systems with orbital periods of about a few days (see Table~\ref{tab:vrotrv}). 

\begin{figure}[t!]
\begin{center}
	\resizebox{\hsize}{!}{\includegraphics{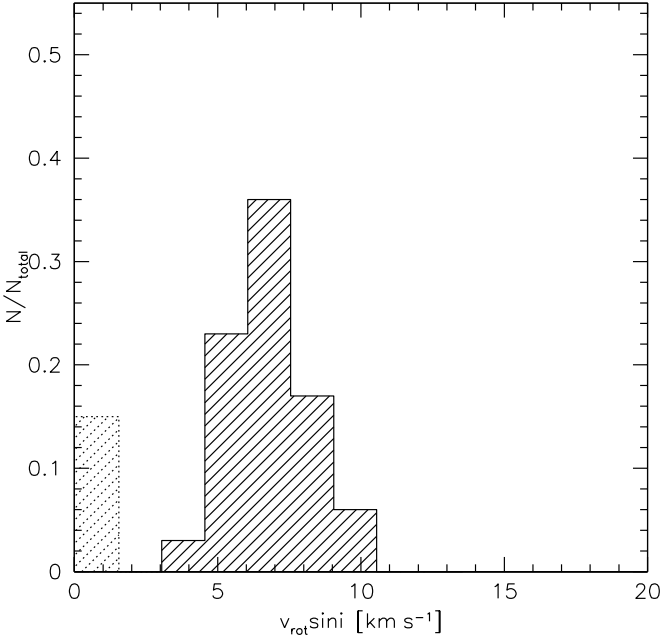}}
	\caption{Distribution of ${v_{\rm rot}\sin\,i}$ for the full sample. Objects with limits below the detection limit have been stacked into the first dotted bin.}
	\label{fig:distriball}
\end{center}
\end{figure}

\begin{figure}[h!]
\begin{center}
	\resizebox{\hsize}{!}{\includegraphics{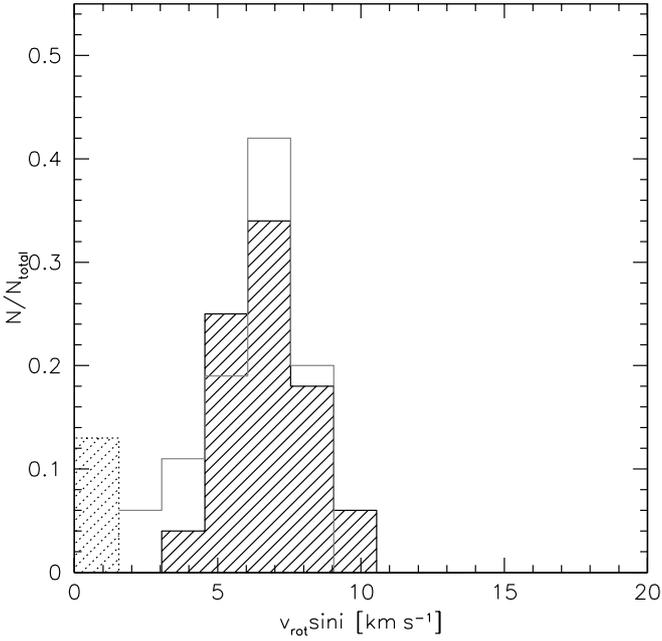}}
	\caption{Distribution of ${v_{\rm rot}\sin\,i}$ for 71 single stars from our sample using the same binning as in Fig.~\ref{fig:distriball}. The solid grey line marks the distribution of ${v_{\rm rot}\sin\,i}$ under the assumption of randomly oriented rotation axes and a constant ${v_{\rm rot}=7.65\,{\rm km\,s^{-1}}}$, which matches the observed distribution very well.}
	\label{fig:distribsingle}
\end{center}
\end{figure}

\begin{figure}[t!]
\begin{center}
	\resizebox{\hsize}{!}{\includegraphics{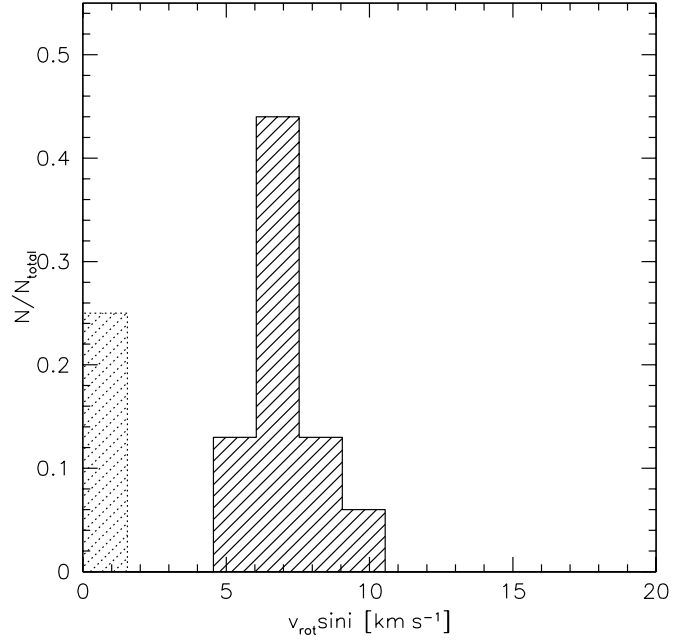}}
	\caption{Distribution of ${v_{\rm rot}\sin\,i}$ for 16 sdBs with companions visible in the spectra using the same binning as in Fig.~\ref{fig:distriball}.}
	\label{fig:distribcomp}
\end{center}
\end{figure}

\begin{figure}[h!]
\begin{center}
	\resizebox{\hsize}{!}{\includegraphics{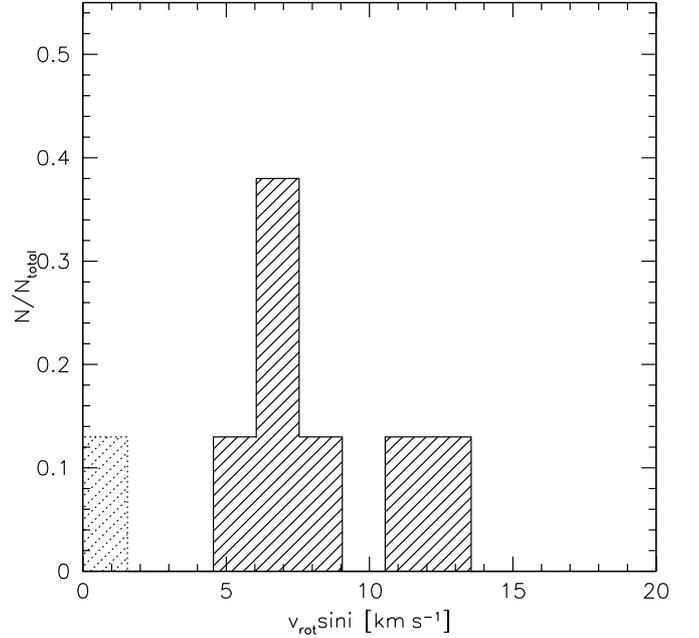}}
	\caption{Distribution of ${v_{\rm rot}\sin\,i}$ for 8 radial velocity variable sdBs with orbital periods exceeding $\simeq1.2\,{\rm d}$ using the same binning as in Fig.~\ref{fig:distriball}.}
	\label{fig:distribrv}
\end{center}
\end{figure}

For eleven stars of our sample upper limits for the projected rotational velocities have already been published (Heber et al. \cite{heber00}; Edelmann et al. \cite{edelmann01}) based on the same spectra as used here (see Table~\ref{tab:vrotlit}). Only for PHL\,932 and PG\,0909$+$276 our measured $v_{\rm rot}\sin{i}$ deviate significantly from the results of Edelmann et al. (\cite{edelmann01}). This is most likely because they used fewer metal lines in their study. 

Przybilla et al. (\cite{przybilla06}) performed an NLTE analysis of Feige\,49 and HD\,205805 using the same FEROS spectra as we do here and derived a ${v_{\rm rot}\sin\,i}$ below the detection limit. Again our measurements are consistent with their results, because they are very close to the detection limit we derived for FEROS spectra of sdBs ($\simeq5\,{\rm km\,s^{-1}}$, see Paper~I). 

\section{Projected rotational velocity distributions \label{sec:distrib}}

The projected rotational velocities of our full sample of 98 stars without radial velocity variations are all low ($<10\,{\rm km\,s^{-1}}$, see Table~\ref{tab:vrot}). Taking into account the uncertainties, one can see that there is no obvious trend with the atmosperic parameters (see Fig.~\ref{fig:tefflogg}). 

Fig.~\ref{fig:distriball} shows the distribution of ${v_{\rm rot}\sin\,i}$ binned to the average measurement error ($1.5\,{\rm km\,s^{-1}}$). Eleven stars that had only fairly weak upper limits of $10\,{\rm km\,s^{-1}}$, were sorted out.
The distribution is very uniform and shows a prominent peak at $6-8\,{\rm km\,s^{-1}}$. Because we can only determine the projected rotation, the true rotational velocities of most stars in the sample should be about $7-8\,{\rm km\,s^{-1}}$. 

\subsection{Single-lined sdBs}

Our sample contains 71 single-lined sdBs, of which the ${v_{\rm rot}\sin\,i}$ could be constrained. Ten stars of which we were only able to derive upper limits of $10\,{\rm km\,s^{-1}}$ were sorted out. Fig.~\ref{fig:distribsingle} shows the ${v_{\rm rot}\sin\,i}$ distribution of this subsample. Most remarkably, the distribution is almost identical to that of the full sample. Adopting a random distribution of inclination angles and a constant ${v_{\rm rot}}$ of $\simeq8\,{\rm km\,s^{-1}}$, the observed ${v_{\rm rot}\sin\,i}$-distribution can indeed be well reproduced (see Fig.~\ref{fig:distriball}). We therefore conclude that most single sdBs in our sample have very similar rotation velocities.

\subsection{Double-lined sdB binaries}

Our sample contains 18 sdBs with visible spectral signatures of cooler main sequence (MS) companions (e.g. Mg\,{\sc i}, Lisker et al. \cite{lisker05}). Again, two stars with upper limits of $10\,{\rm km\,s^{-1}}$ were excluded. 

The orbital periods of these systems are long. Green et al. (\cite{green06}) have argued that such systems should have periods of many months or years. Recently, Deca et al. (\cite{deca12}) were able to determine the orbital period $P\simeq760\,{\rm d}$ of the sdB+K binary PG\,1018$-$047. Similar periods were reported by \O stensen \& van Winckel (\cite{oestensen12}) for eight such binaries. The separations of the components are so wide that tidal interaction is negligible. Main-sequence companions do therefore not affect the rotational properties of the sdB stars in this type of binaries.

The distribution for sdBs with composite spectra is displayed in Fig.~\ref{fig:distribcomp}. Taking into account the much smaller sample size, the result is again similar. We therefore conclude that the rotational properties of sdBs in wide binaries with MS companions are the same as those of single sdBs, although they have probably formed in a very different way (see Sect.~\ref{sec:implications}).

\subsection{Pulsating sdBs}

Two types of sdB pulsators are known. The slow pulsations of the V\,1093\,Her stars (sdBV$_{\rm s}$, Green et al. \cite{green03}) are not expected to influence the line broadening significantly (see Geier et al. \cite{geier10b}). For the short-period pulsators (V\,361\,Hya type, sdBV$_{\rm r}$, Charpinet et al. \cite{charpinet97}; Kilkenny et al. \cite{kilkenny97}) unresolved pulsations can severely affect or even dominate the broadening of the metal lines and therefore fake high ${v_{\rm rot}\sin\,i}$. Telting et al. (\cite{telting08}) showed that this happens in the case of the hybrid pulsator Balloon\,090100001 using the same method as in this work. Unresolved pulsations are also most likely responsible for the high line broadening ($39\,{\rm km\,s^{-1}}$) measured for the strong pulsator PG\,1605+072 (Heber et al. \cite{heber99}, \cite{heber00}).

Our sample contains three known long-period pulsators (PHL\,44, Kilkenny et al. \cite{kilkenny07}; PHL\,457, Blanchette et al. \cite{blanchette08}; LB\,1516, Koen et al. \cite{koen10}) and two short-period ones (KPD\,2109$+$4401, Bill\`{e}res et al. \cite{billeres98}; PG\,1219$+$534, O'Donoghue et al. \cite{odonoghue99}). The ${v_{\rm rot}\sin\,i}$ of KPD\,2109$+$4401 is indeed among the highest of all sdBs in our sample ($10.5\pm1.6\,{\rm km\,s^{-1}}$), but it is unclear if this might not be partly due to unresolved pulsations. Jeffery \& Pollacco (\cite{jeffery00}) measured RV variations of $2\,{\rm km\,s^{-1}}$ for KPD\,2109$+$4401. Taking this into account, the sdBs rotational velocity may be slightly lower than measured. The ${v_{\rm rot}\sin\,i}$ of the other pulsators are not peculiar.

For most stars in our sample it is not clear whether they are pulsators or not, because no light curves of sufficient quality are available. Because only about $5\%$ of all sdBs show pulsations detectable from the ground, one may conclude that the contamination by pulsators should be quite low. Thanks to the extensive photometric surveys for sdB pulsators conducted by Bill\`{e}res et al. (\cite{billeres02}), Randall et al. (\cite{randall06}) and \O stensen et al. (\cite{oestensen10}), we know that 27 stars from our sample do not show short-period pulsations. 

Restricting ourselves to these objects and again excluding those with visible companions, we constructed a "pure" sample of 16 single sdBs, for which the rotational broadening is proven to be disturbed neither by the presence of a companion nor by pulsations. The associated ${v_{\rm rot}\sin\,i}$ distribution does not differ from the other distributions (see Figs.~\ref{fig:distriball}-\ref{fig:distribcomp}). We therefore conclude that unresolved pulsations do not significantly affect our results.

\subsection{Radial velocity variable sdBs}

In Paper~I we showed that the ${v_{\rm rot}\sin\,i}$ distribution of sdBs in close binary systems is strongly affected by the tidal interaction with their companions, but that this influence becomes negligible if the orbital periods of the binaries become longer than $\simeq1.2\,{\rm d}$. It is instructive to have a look at the ${v_{\rm rot}\sin\,i}$-distribution of these long-period radial velocity variable systems. From Paper~I we selected all seven binaries with periods longer than $1.2\,{\rm d}$, for which tidal synchronisation is not established. We added the system LB\,1516, a binary with yet unknown orbital parameters, but for which Edelmann et al. (\cite{edelmann05}) provided a lower limit for the period of the order of days\footnote{TON\,S\,135 was not included because the orbital period of $\simeq4\,{\rm d}$ given in Edelmann et al. (\cite{edelmann05}) is not very  significant and shorter periods cannot be excluded yet.}. 

Fig.~\ref{fig:distribrv} shows the associated distribution. Given the small sample size and although two stars have somewhat higher ${v_{\rm rot}\sin\,i}=10-12\,{\rm km\,s^{-1}}$, the distribution is again very similar to the distributions shown before (see Figs.~\ref{fig:distriball}-\ref{fig:distribcomp}). Subdwarf B stars in close binaries obviously rotate in the same way as single stars or sdBs with visible companions if the orbital period is sufficiently long. 

\section{Comparison with BHB stars \label{sec:bhb}}

\begin{figure}[t!]
\begin{center}
	\resizebox{\hsize}{!}{\includegraphics{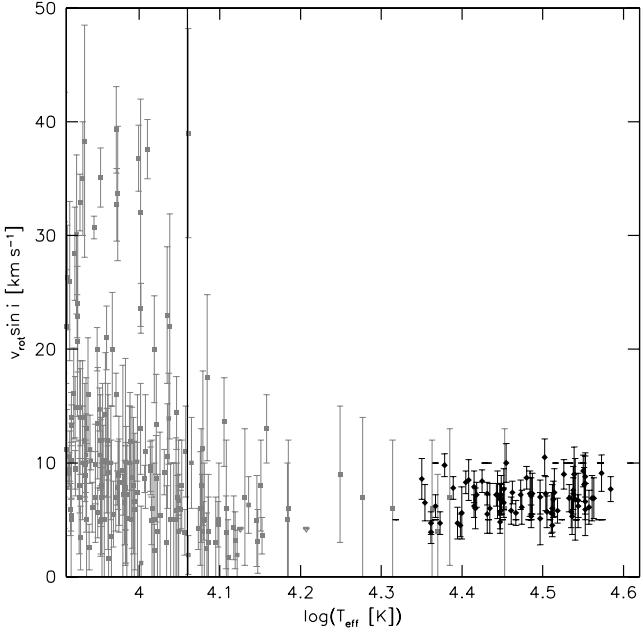}}
	\caption{Projected rotational velocity plotted against effective temperature. The grey squares mark BHB and some sdB stars taken from Peterson et al. (\cite{peterson95}), Behr (\cite{behr03a, behr03b}), Kinman et al. (\cite{kinman00}), and Recio-Blanco et al. (\cite{recio04}). Upper limits are marked with grey triangles. The black diamonds mark the sdBs from our sample. The vertical line marks the jump temperature of $11\,500\,{\rm K}$.}
	\label{fig:vsiniteff}
\end{center}
\end{figure}

Projected rotational velocities of BHB stars have been determined for many globular cluster and field stars (Peterson et al. \cite{peterson95}; Behr \cite{behr03a, behr03b}; Kinman et al. \cite{kinman00}; Recio-Blanco et al. \cite{recio04}). The results are plotted against the effective temperature in Fig.~\ref{fig:vsiniteff}. The characteristic jump in ${v_{\rm rot}\sin\,i}$ at a temperature of about $\simeq11\,500\,{\rm K}$ can be clearly seen. The sdB sequence basically extends the BHB trend to higher temperatures. The ${v_{\rm rot}\sin\,i}$ values remain at the same level as observed in hot BHB stars.

Comparing the ${v_{\rm rot}\sin\,i}$ of BHB and EHB stars, one has to take into account that the radii of both types of horizontal branch stars are quite different, which translates directly into very different angular momenta. While sdBs have surface gravities $\log{g}$ between $5.0$ and $6.0$, the surface gravities of BHB stars range from $\log{g}=3.0$ to $4.0$. The BHB stars with the same rotational velocities as EHB stars have higher angular momenta. Assuming rigid rotation, the same inclination angle of the rotation axis, and the same mass of $\simeq0.5\,M_{\rm \odot}$ for BHB and EHB stars, one can calculate the quantity ${v_{\rm rot}\sin\,i}\times g^{-1/2}$, which is directly proportional to the angular momentum. The surface gravities of the sdBs were taken from the literature (see Sect.~\ref{sec:obs}), those for the BHB stars from Behr (\cite{behr03a, behr03b}) and Kinman et al. (\cite{kinman00}). Since Peterson et al. (\cite{peterson95}) and Recio-Blanco et al. (\cite{recio04}) did not determine surface gravities for their BHB sample, we adopted a $\log{g}$ of $3.0$ for stars with temperatures below $\simeq10\,000\,{\rm K}$ and $3.5$ for the hotter ones as suggested by the results of Behr (\cite{behr03a, behr03b}) and Kinman et al. (\cite{kinman00}).

\begin{figure}[t!]
\begin{center}
	\resizebox{\hsize}{!}{\includegraphics{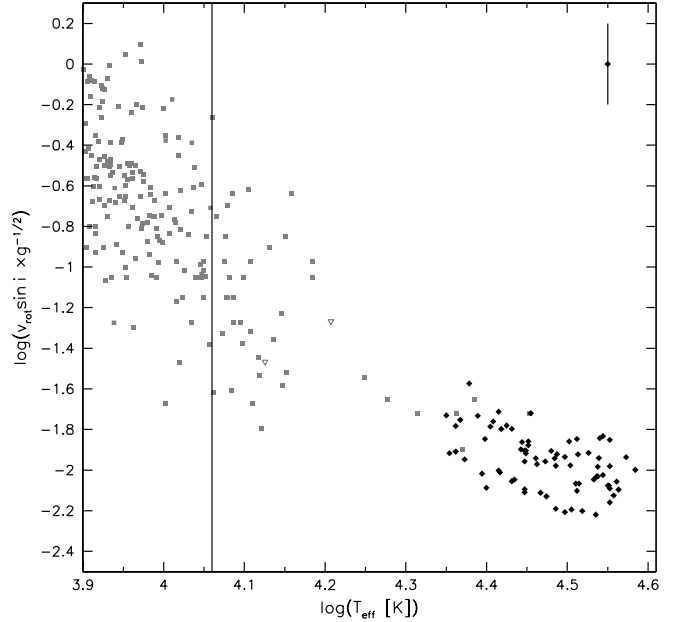}}
	\caption{${v_{\rm rot}\sin\,i}\times g^{-1/2}$ plotted against effective temperature. The grey squares mark BHB and some sdB stars taken from Peterson et al. (\cite{peterson95}), Behr (\cite{behr03a, behr03b}), Kinman et al. (\cite{kinman00}), and Recio-Blanco et al. (\cite{recio04}). Upper limits are marked with grey triangles. The black diamonds mark the sdBs from our sample. The vertical line marks the jump temperature of $11\,500\,{\rm K}$. Typical uncertainties for the sdBs are given in the upper right corner.}
	\label{fig:lteff}
\end{center}
\end{figure}

In Fig.~\ref{fig:lteff} ${v_{\rm rot}\sin\,i}\times g^{-1/2}$ is plotted against $T_{\rm eff}$. The transition between BHB and EHB stars is smooth. Since the progenitors of the EHB stars lost more envelope material on the RGB, the EHB stars are expected to have lower angular momenta than the BHB stars. This is consistent with what can be seen in the Fig.~\ref{fig:lteff}. 

\section{Implications for hot subdwarf formation \label{sec:implications}}

The uniform distribution of low projected rotational velocities in single and wide binary sdBs has consequences for the open question of hot subdwarf formation. As shown in this study, sdBs appear to rotate at low but spectroscopically detectable velocities of $8-10\,{\rm km\,s^{-1}}$. These results are remarkably similar to those derived for their cooler relatives, the BHB stars. Hot subdwarfs are likely formed through binary interaction or merging, which is also accompanied by a transfer of angular momentum. The rotational properties of sdB stars therefore allow one to constrain possible formation scenarios.

\subsection{Uniform rotation of EHB stars and mass loss on the RGB}

The rotational properties of sdBs residing on the EHB are very similar to those of hot BHB stars. The only exception is that the EHB stars obviously lost more envelope in the red-giant phase and therefore retained less angular momentum. How the envelope is lost does not affect the rotational velocities of sdB stars, since the ${v_{\rm rot}\sin\,i}$-distribution of RV variable systems with orbital periods sufficiently long to neglect the tidal influence of the companion (Fig.~\ref{fig:distribrv}) is similar to those of apparently single sdB stars (Fig.~\ref{fig:distribsingle}) and for sdB stars with visible main sequence companions (Fig.~\ref{fig:distribcomp}).

The abundance patterns of sdBs are dominated by diffusion processes very similar to those of the hot BHB stars (Geier et al. \cite{geier10a}). No surface convection zone should be present, and according to the model of Brown (\cite{brown07}) the angular momentum of the outer layers should be low. Stellar winds and magnetic fields may help to slow down the upper layers of the star. However, Unglaub (\cite{unglaub08}) showed that the weak winds predicted for sdB stars are most likely fractionated and are therefore not able to carry away the most abundant elements hydrogen and helium. 

Angular momentum gained or retained from the formation process may also be stored in the stellar core, which may be rapidly rotating. Kawaler \& Hostler (\cite{kawaler05}) proposed such a scenario and suggested an asteroseismic approach to probe the rotation of the inner regions of sdBs. Van Grootel et al. (\cite{vangrootel08}) and Charpinet et al. (\cite{charpinet08}) performed such an analysis for the two short-period sdB pulsators Feige\,48 and PG\,1336$-$018, respectively, and found no deviation from rigid rotation at least in the outer layers of these stars down to about half the stellar radius. But these results may not be representative, because both stars are in close binary systems and are synchronised by the tidal influence of their companions (Geier et al. \cite{geier10b}). The rigid body rotation may have been caused by this effect and may not be a general feature of sdBs. Another setback of these analyses is the problem that p-mode pulsations are not suited to probe the innermost regions of sdBs. In contrast to that, g-mode pulsations reach the stellar core and it should be possible to measure the rotational properties of the whole stellar interior with asteroseismic methods. With the availability of high-precision light curves from the Kepler and CoRoT missions, the analysis of g-mode pulsators became possible and first results have been published by van Grootel et al. (\cite{vangrootel10}) and Charpinet et al. (\cite{charpinet11b}). 

For the RV variable systems CE ejection is the only feasible formation channel. The systems with visible companions may have lost their envelopes via stable RLOF. Very recently, \O stensen et al. (\cite{oestensen12}) and Deca et al. (\cite{deca12}) reported the discovery of sdB+MS binaries with orbital periods up to $\simeq1200\,{\rm d}$, which may have been sufficiently close for mass transfer. 
However, the visible companions to the sdBs may still have been separated by too much for an interaction with the subdwarf progenitors. More detailed binary evolution calculations are needed to solve this problem. Common envelope ejection and stable RLOF form similar sdB stars, because in both cases the hydrogen envelope is removed and the helium burning should start under similar conditions. It would therefore not be surprising if their ${v_{\rm rot}\sin\,i}$-distributions were to look similar.

\subsection{Where are the He-WD merger products?}

The ${v_{\rm rot}\sin\,i}$-distribution of the single sdB stars (Fig.~\ref{fig:distribsingle}) is particularly hard to understand in the context of the WD merger scenario. If a certain fraction or even all of the apparently single sdBs would have been formed in this way, one would not expect a ${v_{\rm rot}\sin\,i}$-distribution that resembles that of the post-CE or post-RLOF sdBs. Gourgouliatos \& Jeffery (\cite{gourgouliatos06}) showed that the merger product of two WDs would rotate faster than break-up velocity, if angular momentum were conserved. These authors concluded that angular momentum must be lost during the merger process. One way to lose angular momentum are stellar winds and magnetic fields. Another explanation may be the interaction with the accretion disc during the merger. If the less massive object is disrupted, it should form an accretion disc around the more massive component. The WD can only gain mass if angular momentum is transported outward in the disc. This process is expected to spin down the merger product (Gourgouliatos \& Jeffery \cite{gourgouliatos06}). According to a model proposed by Podsiadlowski (priv. comm.), the merger is accompanied by a series of outbursts caused by the ignition of helium. These flashes remove angular momentum from the merged remnant and should slow it down to rotational velocities of less than $20\,{\rm km\,s^{-1}}$. 

However, even if it is possible to slow down the merged remnant of two He-WDs, it is very unlikely that the merger pro\-ducts would have a ${v_{\rm rot}\sin{i}}$-distribution almost identical to sdBs, of which we know that they were formed via CE-ejection or maybe stable RLOF. This would require an extreme fine-tuning of parameters, unless there is an as yet unknown mechanism at work, which leads to uniform rotation of the radiative, diffusion-dominated atmospheres. It is therefore questionable whether our sample contains stars that were formed by an He-WD merger or a CE-merger event. If this is not the case and because of the size of our sample, it would be safe to conclude that the merger channel does not contribute significantly to the observed population of single hydrogen-rich sdO/Bs in contrast to the models of Han et al. (\cite{han02}, \cite{han03}). 

This conclusion is consistent with the most recent results by Fontaine et al. (\cite{fontaine12}), who studied the empirical mass distribution of sdB stars derived from eclipsing binary systems and asteroseismic analyses. The lack of sdB stars more massive than $\simeq0.5\,M_{\odot}$, which would be the outcome of the merger channel, led to the conclusion that mergers are less frequent in the formation process of isolated sdB stars than predicted by theory.

The only known single and fast rotating hot subdwarf star EC\,22081$-$1916 (Geier et al. \cite{geier11a}) may be the rare outcome of a CE merger event as suggested by Politano et al. (\cite{politano08}). It is unique among $\simeq100$ sdBs of our sample.

Possible candidates for WD-merger products are the helium rich sdOs (He-sdOs, Str\"oer at al. \cite{stroeer07}), since Hirsch et al. (\cite{hirsch09}) measured ${v_{\rm rot}\sin\,i}$ values of $20-30\,{\rm km\,s^{-1}}$ for some of those stars. Although their velocities are not particularly high, they are significantly different from the typical ${v_{\rm rot}\sin\,i}$ of sdBs. However, while the He-sdOs were first  considered as single stars (Napiwotzki et al. \cite{napiwotzki08}), evidence grows that a fraction of them resides in close binaries (Green et al. \cite{green08}; Geier et al. \cite{geier11b}). At least those He-sdOs could not have been formed by a He-WD merger. 

\subsection{Alternative formation scenarios}

Because the canonical binary scenario for sdB formation, which rests on the three pillars CE ejection, stable RLOF and He-WD merger, turned out to be very successful not only in explaining the properties of sdBs in the field (Han et al. \cite{han02}, \cite{han03}), but also in globular clusters (Han \cite{han08}) and the UV-upturn phenomenon in old galaxies (Han et al. \cite{han07}), the possible lack of merger candidates poses a problem. 

Alternative formation scenarios such as CE ejection triggered by substellar companions (Soker \cite{soker98}; Bear \& Soker \cite{bear12}) may be responsible for the formation of apparently single sdBs. Evidence grows that such objects are quite common around sdB stars (e.g. Silvotti et al. \cite{silvotti07}; Geier et al. \cite{geier11c}; Charpinet et al. \cite{charpinet11a}). In the light of the results presented here and other recent observational evidence, the conclusion has to be drawn that the question of sdB formation is still far from settled. 

\section{Connection to white dwarfs \label{sec:wd}}

Owing to their thin hydrogen envelopes, hot subdwarf stars will not evolve to the asymptotic giant branch (AGB-manqu\'e, Dorman et al. \cite{dorman93}). After about $100\,{\rm Myr}$ of core He-burning on the EHB and a shorter episode of He-shell burning, these objects will join the WD cooling sequence. 

The rotational properties of single WDs are difficult to determine. Owing to the high pressure in the dense WD atmospheres, the spectral lines of WDs are strongly broadened and hence do not appear to be suitable to measure ${v_{\rm rot}\sin{i}}$. However, the H${\rm_\alpha}$ line often displays a sharp line core, which is caused by NLTE effects. In a small fraction of the WD-population metal lines are visible. However, excellent high-resolution spectra are necessary to constrain the projected rotational velocity (Berger et al. \cite{berger05}).

The derived upper limits ($\simeq10-50\,{\rm km\,s^{-1}}$) are consistent with the much lower rotational velocities of pulsating WDs derived with asteroseismic methods ($\simeq0.2-3.5\,{\rm km\,s^{-1}}$, Kawaler \cite{kawaler03}). Most single WDs are therefore obviously rather slow rotators. The reason for this is most likely a significant loss of mass and angular momentum due to stellar winds and thermal pulses in the AGB-phase, as has been shown by Charpinet et al. (\cite{charpinet09}).

The properties of WDs evolved from sdB progenitors on the other hand should be very different. Since the hot subdwarfs bypass the AGB-phase, both their masses and their angular momenta are expected to remain more or less constant when evolving to become WDs. 

The average mass of these sdB remnants ($\simeq0.47\,M_{\rm \odot}$) is expected to be significantly lower than the average mass of normal WDs ($\simeq0.6\,M_{\rm \odot}$). But more importantly, the rotational velocities of these WDs must be very high. We have shown that single sdBs have small, but still detectable ${v_{\rm rot}\sin{i}}$. Assuming rigid rotation and conservation of mass and angular momentum, the rotational velocity at the surface scales with the stellar radius. Because the radius decreases by a factor of about $10$, the rotational velocity should increase by a factor of about $100$. Assuming an average ${v_{\rm rot}\simeq8\,{\rm km\,s^{-1}}}$ for single sdBs, WDs evolved through an EHB-phase should therefore have an average ${v_{\rm rot}\simeq800\,{\rm km\,s^{-1}}}$. Because about $1\%$ of all WDs are expected to have evolved through an EHB-phase, we expect a similar fraction of extremely fast rotating, low-mass WDs. These high ${v_{\rm rot}\sin{i}}$-values should be easily detectable even in medium-resolution spectra. The sample of WDs with observed spectra from the Sloan Digital Sky Survey (Eisenstein et al. \cite{eisenstein06}) for example should contain more than $100$ of these objects. 

\section{Summary \label{sec:summary}}

We extended a project to derive the rotational properties of sdB stars and determined the projected rotational velocities of 105 sdB stars by measuring the broadening of metal lines using high-resolution spectra. All stars in our sample have low ${v_{\rm rot}\sin{i}}<10\,{\rm km\,s^{-1}}$. For $\simeq75\%$ of the sample we were able to determine significant rotation. The distribution of projected rotational velocities is consistent with an average rotation of $\simeq8\,{\rm km\,s^{-1}}$ for the sample. Furthermore, the $v_{\rm rot}\sin{i}$-distributions of single sdBs, hot subdwarfs with main sequence companions vi\-sible in the spectra and close binary systems with periods exceeding $1.2\,{\rm d}$ are similar. The BHB and EHB stars are related in terms of surface rotation and angular momentum. Hot BHBs with diffusion-dominated atmospheres are slow rotators like the EHB stars, which lost more envelope and therefore angular momentum on the RGB. The uniform rotation distributions of single and wide binary sdBs pose a challenge to our understanding of hot subdwarf formation. Especially the high fraction of He-WD mergers predicted by theory seems to be inconsistent with our results. We predict that the evolutionary channel of single sdB stars gives birth to a small population of rapidly rotating WDs with masses lower than average.

\begin{table*}[t!]
\caption{Projected rotational velocities of single sdBs and sdBs with visible companions.} 
\label{tab:vrot}
\begin{center}
\begin{tabular}{llllllll}
\hline
\noalign{\smallskip}
 System & $T_{\rm eff}$ & $m_{B/V}$ & S/N & seeing &  $N_{\rm lines}$ &  ${v_{\rm rot}\,\sin\,i}$ & Instrument \\ 
& [K] & [mag] &  & [arcsec] &  & [${\rm km\,s^{-1}}$]  \\
\noalign{\smallskip}
\hline
\noalign{\smallskip}
HE\,0151$-$3919 & 20\,800 & 14.3$^{\rm B}$ &  66 & 1.06 & 27 &  $<5.0$ & UVES  \\
EC\,21494$-$7018 & 22\,400 & 11.2$^{\rm V}$ & 85 &  & 16 &  8.6 $\pm$ 1.8 & FEROS   \\
EC\,15103$-$1557 & 22\,600 & 12.9$^{\rm V}$ & 163 &  &  8 & 6.5 $\pm$ 1.6 & FEROS   \\
HD\,4539 & 23\,000 & 10.1$^{\rm B}$ & 112 &  & 21 &  3.9 $\pm$ 1.0 & FEROS   \\
EC\,11349$-$2753 & 23\,000 & 12.5$^{\rm B}$ & 185 &  & 49 & 4.7 $\pm$ 1.0 & FEROS   \\
EC\,14345$-$1729 & 23\,300 & 13.1$^{\rm V}$ & 117 &  & 40 & 6.2 $\pm$ 1.0 & FEROS   \\
HE\,0539$-$4246 & 23\,300 & 14.5$^{\rm B}$ &  40 & 0.87 & 19 &  $<10.0$ & UVES   \\
HE\,2307$-$0340$^{\rm no}$ & 23\,300 & 15.8$^{\rm B}$ &  61 & 0.89 & 17 &  $<5.0$ & UVES   \\
PG\,1432$+$004$^{\rm nr}$ & 23\,600 & 12.0$^{\rm B}$ & 170 &  & 13 & 4.7 $\pm$ 1.0 & FEROS   \\
EC\,19563$-$7205$^{\rm c}$ & 23\,900 & 12.8$^{\rm B}$ & 85 &  & 34 & 9.8 $\pm$ 1.0 & FEROS   \\
EC\,20106$-$5248 & 24\,500 & 12.6$^{\rm V}$ & 114 &  & 47 &  7.8 $\pm$ 1.0 & FEROS   \\
BD$+$48$^{\circ}$\,2721 & 24\,800 & 10.5$^{\rm B}$ & 326 &  & 10 & 4.7 $\pm$ 1.4 & FOCES \\
HD\,205805 & 25\,000 & 9.9$^{\rm B}$ & 255 &  & 20 &  4.5 $\pm$ 1.0 & FEROS   \\
HE\,0321$-$0918$^{\rm no}$ & 25\,100 & 14.7$^{\rm B}$ &  37 & 1.22 &  7 &  5.6 $\pm$ 2.3 & UVES   \\
PG\,1653$+$131 & 25\,400 & 14.1$^{\rm B}$  & 68 &  & 32 & 8.3 $\pm$ 1.0 & FEROS   \\
HE\,2237$+$0150 & 25\,600 & 15.8$^{\rm B}$  &  40 & 0.78 & 11 &  8.5 $\pm$ 1.8 & UVES   \\
PG\,0342$+$026 & 26\,000 & 11.1$^{\rm B}$  & 190 &  & 54 & 6.2 $\pm$ 1.0 & FEROS   \\
PG\,2122$+$157$^{\rm c}$ & 26\,000 & 15.0$^{\rm B}$  & 67 & 0.78 & 13 & 7.9 $\pm$ 1.4  & UVES   \\
GD\,108 & 26\,100 & 13.3$^{\rm B}$  & 97 &  & 6 & 6.0 $\pm$ 1.8 & FEROS   \\
Feige\,65 & 26\,200 & 11.8$^{\rm B}$  & 150 &  & 18 & 7.2 $\pm$ 1.1 & FOCES   \\
PHL\,44$^{\rm l}$ & 26\,600 & 13.0$^{\rm B}$  & 85 &  & 31 & 8.4 $\pm$ 1.0 & FEROS   \\
HE\,0513$-$2354 & 26\,800 & 15.8$^{\rm B}$  &  21 & 0.99 & 18 &  $<10.0$ & UVES   \\
HE\,0135$-$6150 & 27\,000 & 16.3$^{\rm B}$  &  37 & 0.71 & 13 &  5.5 $\pm$ 1.7 & UVES   \\
SB\,815 & 27\,000 & 10.6$^{\rm B}$  & 85 &  & 48 & 7.3 $\pm$ 1.0 & FEROS   \\
HE\,2201$-$0001 & 27\,100 & 16.0$^{\rm B}$  &  35 & 1.10 & 28 &  $<5.0$ & UVES   \\
PG\,2205$+$023 & 27\,100 & 12.9$^{\rm B}$  & 36 &  & 9 & $<10.0$ & FEROS   \\
PG\,2314$+$076$^{\rm nb}$ & 27\,200 & 13.9$^{\rm B}$  & 71 &  & 6 & 6.0 $\pm$ 2.2 & FEROS   \\
SB\,485 & 27\,700 & 13.0$^{\rm B}$  & 112 & 0.71 & 24 & 7.2 $\pm$ 1.0  & UVES   \\
KUV\,01542$-$0710$^{\rm c}$ & 27\,800 & 16.3$^{\rm B}$  & 58 & 0.92 & 8 & 7.2 $\pm$ 2.1  & UVES   \\
HE\,2156$-$3927$^{\rm c}$ & 28\,000 & 14.1$^{\rm B}$  &  62 & 0.61 & 16 &  7.0 $\pm$ 1.2 & UVES   \\
EC\,03591$-$3232 & 28\,000 & 11.2$^{\rm V}$ & 131 &  & 34 & 4.8 $\pm$ 1.0 & FEROS   \\
EC\,12234$-$2607 & 28\,000 & 13.8$^{\rm B}$ & 60 &  & 19 & 6.8 $\pm$ 1.4 & FEROS   \\
PG\,2349$+$002 & 28\,000 & 12.0$^{\rm B}$ & 68 &  & 11 & 5.7 $\pm$ 1.5 & FEROS   \\
HE\,2322$-$0617$^{\rm c,no}$ & 28\,100 & 15.7$^{\rm B}$ &  62 & 0.70 & 15 &  6.8 $\pm$ 1.3 & UVES   \\
PG\,0258$+$184$^{\rm c,no}$ & 28\,100 & 15.2$^{\rm B}$ & 48 & 0.99 & 12 & 7.2 $\pm$ 1.7  & UVES   \\
HE\,0136$-$2758$^{\rm no}$ & 28\,200 & 16.2$^{\rm B}$ &  29 & 1.20 & 27 &  $<5.0$ & UVES   \\
HE\,0016$+$0044$^{\rm no}$ & 28\,300 & 13.1$^{\rm B}$ &  58 & 0.67 & 14 &  6.5 $\pm$ 1.3 & UVES   \\
PG\,1549$-$001$^{\rm no}$ & 28\,300 & 14.8$^{\rm B}$ & 45 & 1.16 & 20 & 5.6 $\pm$ 1.1  & UVES   \\
HE\,2349$-$3135 & 28\,500 & 15.6$^{\rm B}$ &  53 & 1.13 & 13 &  10.0 $\pm$ 1.7 & UVES   \\
EC\,01120$-$5259 & 28\,900 & 13.5$^{\rm V}$ & 73 &  & 19 & 5.8 $\pm$ 1.2 & FEROS   \\
HE\,0007$-$2212$^{\rm no}$ & 29\,000 & 14.8$^{\rm B}$ &  53 & 0.64 & 21 &  7.4 $\pm$ 1.0 & UVES   \\
LB\,275$^{*}$ & 29\,300 & 14.9$^{\rm B}$ & 48 & 1.16 & 20 & 5.6 $\pm$ 1.1  & UVES   \\
EC\,03263$-$6403 & 29\,300 & 13.2$^{\rm V}$ & 32 &  & 40 & $<5.0$ & FEROS   \\
HE\,1254$-$1540$^{\rm c,no}$ & 29\,700 & 15.2$^{\rm B}$ &  54 & 0.75 & 20 &  7.2 $\pm$ 1.3 & UVES   \\
PG\,1303$+$097 & 29\,800 & 14.3$^{\rm B}$ & 51 &  & 18 & 6.1 $\pm$ 1.5 & FEROS   \\
HE\,2222$-$3738 & 30\,200 & 14.2$^{\rm B}$ &  61 & 0.83 & 28 &  8.7 $\pm$ 1.0 & UVES   \\
HE\,2238$-$1455 & 30\,400 & 16.0$^{\rm B}$ &  48 & 0.80 & 14 &  $<5.0$ & UVES   \\
EC\,03470$-$5039 & 30\,500 & 13.6$^{\rm V}$ & 53 &  &  9 & 7.3 $\pm$ 2.0 & FEROS   \\
Feige\,38 & 30\,600 & 12.8$^{\rm B}$ & 148 &  & 34 & 5.3 $\pm$ 1.0 & FEROS   \\
HE\,1038$-$2326$^{\rm c}$ & 30\,600 & 15.8$^{\rm B}$ &  34 & 1.27 & 28 &  $<5.0$ & UVES   \\
PG\,1710$+$490 & 30\,600 & 12.1$^{\rm B}$ & 80 &  & 11 & 7.1 $\pm$ 1.6 & FOCES   \\
HE\,0447$-$3654 & 30\,700 & 14.6$^{\rm V}$ & 44 &  & 11 &  7.3 $\pm$ 1.8 & FEROS   \\
EC\,14248$-$2647 & 31\,400 & 12.0$^{\rm V}$ & 104 &  & 14 & 7.0 $\pm$ 1.5 & FEROS   \\
HE\,0207$+$0030$^{\rm no}$ & 31\,400 & 14.7$^{\rm B}$ &  27 & 1.30 &  7 &  5.1 $\pm$ 2.3 & UVES   \\
KPD\,2109$+$4401$^{\rm s}$ & 31\,800 & 13.2$^{\rm B}$ & 136 &  & 9 & 10.5 $\pm$ 1.6 & HIRES   \\
EC\,02542$-$3019 & 31\,900 & 12.8$^{\rm B}$ & 65 &  & 13 & 7.3 $\pm$ 1.5 & FEROS  \\
$[$CW83$]$\,1758$+$36$^{\rm nb}$ & 32\,000 & 11.1$^{\rm B}$ & 110 &  & 5 & 5.7 $\pm$ 1.4 & FOCES  \\
TON\,S\,155$^{\rm c}$ & 32\,300 & 14.9$^{\rm B}$ & 35 & 0.85 & 14 & $<5.0$  & UVES  \\
EC\,21043$-$4017 & 32\,400 & 13.1$^{\rm V}$ & 65 &  &  8 &  5.6 $\pm$ 1.8 & FEROS   \\
EC\,20229$-$3716 & 32\,500 & 11.4$^{\rm V}$ & 153 &  & 29 &  4.5 $\pm$ 1.0 & FEROS   \\
HS\,2125$+$1105$^{\rm c}$ & 32\,500 & 16.4$^{\rm B}$ &  29 & 0.80 & 8 &  6.0 $\pm$ 2.4 & UVES   \\
HE\,1221$-$2618$^{\rm c}$ & 32\,600 & 14.9$^{\rm B}$ &  35 & 1.06 & 11 &  6.8 $\pm$ 1.6 & UVES   \\
HS\,2033$+$0821$^{\rm no}$ & 32\,700 & 14.4$^{\rm B}$ &  43 & 1.14 & 37 &  $<5.0$ & UVES   \\
HE\,0415$-$2417$^{\rm no}$ & 32\,800 & 16.2$^{\rm B}$ &  34 & 0.83 & 10 &  $<10.0$ & UVES   \\
EC\,05479$-$5818 & 33\,000 & 13.1$^{\rm V}$ & 81 &  & 20 & 5.8 $\pm$ 1.1 & FEROS   \\
HE\,1200$-$0931$^{\rm c,no}$ & 33\,400 & 16.2$^{\rm B}$ &  30 & 0.86 & 12 &  $<5.0$ & UVES   \\
\hline
\\
\end{tabular}
\end{center}
\end{table*}

\begin{table*}[t!]
\begin{center}
\begin{tabular}{llllllll}
\hline
\noalign{\smallskip}
 System & $T_{\rm eff}$ & $m_{B}$ & S/N & seeing &  $N_{\rm lines}$ &  ${v_{\rm rot}\,\sin\,i}$ & Instrument \\ 
& [K] & [mag] &  & [arcsec] &  & [${\rm km\,s^{-1}}$]  \\
\noalign{\smallskip}
\hline
\noalign{\smallskip}
PHL\,932 & 33\,600 & 12.0$^{\rm B}$ & 102 & 1.10 & 12 & 9.0 $\pm$ 1.3 & UVES   \\
HE\,1422$-$1851$^{\rm c,no}$ & 33\,900 & 16.3$^{\rm B}$ &  14 & 0.58 & 10 &  $<10.0$ & UVES   \\
PHL\,555 & 34\,100 & 13.8$^{\rm B}$ & 56 & 0.88 & 17 & 6.9 $\pm$ 1.2  & UVES   \\
HE\,1419$-$1205$^{\rm c}$ & 34\,200 & 16.2$^{\rm B}$ &  28 & 0.69 & 16 &  $<10.0$ & UVES   \\
PG\,1219$+$534$^{\rm s}$ & 34\,300 & 12.4$^{\rm B}$ & 140 &  & 11 & 5.7 $\pm$ 1.4 & HIRES   \\
HS\,2216$+$1833$^{\rm c}$ & 34\,400 & 13.8$^{\rm B}$ &  54 & 0.90 & 11 &  5.3 $\pm$ 1.6 & UVES   \\
HE\,1050$-$0630$^{\rm no}$ & 34\,500 & 14.0$^{\rm B}$ &  59 & 1.20 & 28 &  7.3 $\pm$ 1.4 & UVES   \\
HE\,1519$-$0708$^{\rm no}$ & 34\,500 & 15.6$^{\rm B}$ &  20 & 0.84 & 8 &  9.0 $\pm$ 2.4 & UVES   \\
HE\,1450$-$0957 & 34\,600 & 15.1$^{\rm B}$ &  32 & 0.71 & 6 &  9.0 $\pm$ 2.4 & UVES   \\
EC\,13047$-$3049 & 34\,700 & 12.8$^{\rm V}$ & 68 &  &  5 & 6.8 $\pm$ 3.6 & FEROS   \\
HS\,1710$+$1614$^{\rm no}$ & 34\,800 & 15.7$^{\rm B}$ &  38 & 1.30 & 13 &  $<5.0$ & UVES   \\
PHL\,334 & 34\,800 & 12.5$^{\rm B}$ & 87 &  & 13 & $<5.0$ & FEROS   \\
Feige\,49 & 35\,000 & 13.2$^{\rm B}$ & 119 &  & 40 & 6.2 $\pm$ 1.0 & FEROS   \\
HE\,2151$-$1001$^{\rm s}$ & 35\,000 & 15.6$^{\rm B}$ &  42 & 0.66 & 6 &  6.7 $\pm$ 2.4 & UVES   \\
PG\,0909$+$164$^{\rm s}$ & 35\,300 & 13.9$^{\rm B}$ & 52 &  & 4 & $<10.0$ & FEROS   \\
HE\,1021$-$0255$^{\rm no}$ & 35\,500 & 15.3$^{\rm B}$ &  40 & 1.61 & 11 &  $<10.0$ & UVES   \\
PG\,0909$+$276$^{\rm nb}$ & 35\,500 & 13.9$^{\rm B}$ & 82 &  & 13 & 9.3 $\pm$ 1.4 & FOCES   \\
HE\,0101$-$2707 & 35\,600 & 15.0$^{\rm B}$ &  67 & 0.85 & 12 &  8.1 $\pm$ 1.5 & UVES   \\
EC\,03408$-$1315 & 35\,700 & 13.6$^{\rm V}$ & 66 &  & 11 & 8.8 $\pm$ 1.8 & FEROS   \\
HE\,1352$-$1827$^{\rm c}$ & 35\,700 & 16.2$^{\rm B}$ &  24 & 0.85 & 5 &  8.2 $\pm$ 2.7 & UVES   \\
PG\,1207$-$032$^{\rm no}$ & 35\,700 & 13.1$^{\rm B}$ & 50 & 0.64 & 9 & 6.6 $\pm$ 1.6  & UVES   \\
HE\,0019$-$5545 & 35\,700 & 15.8$^{\rm B}$ &  38 & 0.76 &  7 &  5.9 $\pm$ 2.3 & UVES   \\
GD\,619 & 36\,100 & 13.9$^{\rm B}$ & 96 & 0.81 & 10 & 6.1 $\pm$ 1.5  & UVES   \\
HE\,1441$-$0558$^{\rm c,no}$ & 36\,400 & 14.4$^{\rm B}$ &  30 & 0.70 & 8 &  6.9 $\pm$ 2.0 & UVES   \\
HE\,0123$-$3330 & 36\,600 & 15.2$^{\rm B}$ &  48 & 0.66 &  8 &  6.9 $\pm$ 1.8 & UVES   \\
PG\,1505$+$074 & 37\,100 & 12.2$^{\rm B}$ & 153 &  & 4 & $<5.0$ & FEROS   \\
HE\,1407$+$0033$^{\rm no}$ & 37\,300 & 15.5$^{\rm B}$ &  35 & 0.72 & 9 &   $<10.0$ & UVES   \\
PG\,1616$+$144$^{\rm nb}$ & 37\,300 & 13.5$^{\rm B}$ & 44 &  & 4 & $<10.0$ & FEROS   \\
EC\,00042$-$2737$^{\rm c}$ & 37\,500 & 13.9$^{\rm B}$ & 37 &  & 9 & $<10.0$ & FEROS  \\
PHL\,1548 & 37\,400 & 12.5$^{\rm B}$ & 90 &  & 10 & 9.1 $\pm$ 1.6 & FEROS   \\
PB\,5333$^{\rm nb}$ & 40\,600 & 12.5$^{\rm B}$ & 66 &  & 2 &  $<10.0$ & FEROS   \\
$[$CW83$]$\,0512$-$08 & 38\,400 & 11.3$^{\rm B}$ & 124 &  & 14 & 7.7 $\pm$ 1.1 & FEROS   \\
\hline
\\
\end{tabular}
\tablefoot{The average seeing is only given if the spectra were obtained with a wide 
slit in the course of the SPY survey. In all other cases the seeing should not 
influence the measurements. $^{\rm c}$Main sequence companion visible in the spectrum (Lisker et al. \cite{lisker05}). $^{\rm s}$Pulsating subdwarf of V\,361\,Hya type. $^{\rm l}$Pulsating subdwarf of V\,1093\,Her type. No short-period pulsations have been detected either by $^{\rm nb}$Bill\`{e}res et al. (\cite{billeres02}), $^{\rm nr}$Randall et al. (\cite{randall06}) or $^{\rm no}$\O stensen et al. (\cite{oestensen10}). $^{*}$Misidentified as CBS\,275 in Lisker et al. (\cite{lisker05}).}
\end{center}
\end{table*}

\begin{table*}[t!]
\caption{Projected rotational velocities of radial velocity variable sdBs.} 
\label{tab:vrotrv}
\begin{center}
\begin{tabular}{lllllll}
\hline
\noalign{\smallskip}
 System & $T_{\rm eff}$ & $m_{B/V}$ & S/N & $N_{\rm lines}$ &  ${v_{\rm rot}\,\sin\,i}$ & Instrument\\ 
& [K] & [mag] &  &  & [${\rm km\,s^{-1}}$] & \\
\noalign{\smallskip}
\hline
\noalign{\smallskip}
TON\,S\,135        & 25\,000 & 13.1$^{\rm B}$ & 45 & 35 & 6.4 $\pm$ 1.0 & FEROS \\
LB\,1516$^{\rm l}$ & 25\,200 & 12.7$^{\rm B}$ & 58 & 23 & 6.0 $\pm$ 1.3 & FEROS \\
PHL\,457$^{\rm l}$ & 26\,500 & 13.0$^{\rm B}$ & 59 & 47 & 6.1 $\pm$ 1.0 & FEROS \\
EC\,14338$-$1445 & 27\,700   & 13.5$^{\rm V}$     & 71 & 39 & 8.9 $\pm$ 1.0 & FEROS \\
PG\,1725$+$252  & 28\,900    & 11.5$^{\rm B}$ & 45 & 11 & 7.4 $\pm$ 1.1 & HRS \\
PG\,1519$+$640 & 30\,300 & 12.1$^{\rm B}$ & 104 & 11 & 9.4 $\pm$ 1.4 & FOCES \\
PG\,2151$+$100 & 32\,700 & 12.9$^{\rm B}$ &  69 & 9 & 9.0 $\pm$ 1.7 & FEROS \\
\hline
\\
\end{tabular}
\tablefoot{$^{\rm l}$Pulsating subdwarf of V\,1093\,Her type.}
\end{center}
\end{table*}

\begin{table*}[t!]
\caption{Comparison with literature.} 
\label{tab:vrotlit}
\begin{center}
\begin{tabular}{lrrl}
\hline
\noalign{\smallskip}
System & This work & Literature & Reference \\
 & ${v_{\rm rot}\,\sin\,i}$ & ${v_{\rm rot}\,\sin\,i}$ & \\ 
& [${\rm km\,s^{-1}}$] & [${\rm km\,s^{-1}}$] & \\
\noalign{\smallskip}
\hline
\noalign{\smallskip}
KPD\,2109$+$4401 & $10.5\pm1.6$ & $<10.0$ & Heber \\
PG\,1219$+$534 & $5.7\pm1.4$ & $<10.0$ &  et al. (\cite{heber00}) \\
\noalign{\smallskip}
\hline
\noalign{\smallskip}
BD$+$48$^{\circ}$\,2721 & $4.7\pm1.4$ & $<5.0$ & Edelmann \\
Feige\,65               & $7.2\pm1.1$ & $<5.0$ & et al. (\cite{edelmann01}) \\
HD\,205805              & $4.5\pm1.0$ & $<5.0$ & \\
HD\,4539                & $3.9\pm1.0$ & $<5.0$ & \\
LB\,1516                & $6.0\pm1.3$ & $<5.0$ & \\
PG\,0342$+$026          & $6.2\pm1.0$ & $<5.0$ & \\
PG\,0909$+$276          & $9.3\pm1.4$ & $<5.0$ & \\
PHL\,932                & $9.0\pm1.3$ & $<5.0$ & \\
\noalign{\smallskip}
\hline
\noalign{\smallskip}
Feige\,49               & $6.2\pm1.0$ & $0.0^{*}$ & Przybilla \\
HD\,205805              & $4.5\pm1.0$ & $0.0^{*}$ & et al. (\cite{przybilla06}) \\
\noalign{\smallskip}
\hline
\noalign{\smallskip}
\end{tabular}
\tablefoot{$^{*}$Adopted value for line fits is below the detection limit.}
\end{center}
\end{table*}

\begin{acknowledgements}

S. G. was supported by the Deutsche Forschungsgemeinschaft under grant 
He~1354/49-1. The authors thank N. Reid, R. Napiwotzki, L. Morales-Rueda and H. Edelmann for providing their data. 
Furthermore, we would like to thank the referee G. Fontaine for his comments and suggestions.

\end{acknowledgements}


\begin{thebibliography}{}

\bibitem[2012]{bear12}
Bear, E., \& Soker, N. 2012, ApJ, 749, L14

\bibitem[2003a]{behr03a}
Behr, B. B. 2003, ApJS, 149, 67

\bibitem[2003b]{behr03b}
Behr, B. B. 2003, ApJS, 149, 101

\bibitem[2000a]{behr00a}
Behr, B. B., Cohen, J. G., \& McCarthy, J. K. 2000a, ApJ, 531, L57

\bibitem[2000b]{behr00b}
Behr, B. B., Djorgovski, S. G., Cohen, J. G., et al. 2000b, ApJ, 528, 849

\bibitem[2005]{berger05}
Berger, L., Koester, D., Napiwotzki, R., Reid, I. N., \& Zuckerman, B. 2005, A\&A, 444, 565

\bibitem[1998]{billeres98}
Bill\`{e}res, M., Fontaine, G., Brassard, P., et al. 1998, ApJ, 494, L75

\bibitem[2002]{billeres02}
Bill\`{e}res, M., Fontaine, G., Brassard, P., \& Liebert, J. 2002, ApJ, 578, 515

\bibitem[2008]{blanchette08}
Blanchette, J.-P., Chayer, P., Wesemael, F., et al. 2008, ApJ, 678, 1329

\bibitem[2007]{brown07}
Brown, D. 2007, PhD thesis, Liverpool John Moores University

\bibitem[2008]{brown08}
Brown, D., Salaris, M., Cassisi, S., \& Pietrinferni, A. 2008, Memorie della Societa Astronomica Italiana, 79, 579

\bibitem[1999]{caloi99}
Caloi, V. 1999, A\&A, 343, 904

\bibitem[2008]{carney08}
Carney, B. W., Gray, D. F., Yong, D., et al. 2008, ApJ, 135, 892

\bibitem[2003]{carney03}
Carney, B. W., Latham, D. W., Stefanik, R. P., Laird, J. B., \& Morse, J. A. 2003, AJ, 125, 293

\bibitem[1997]{charpinet97}
Charpinet, S., Fontaine, G., Brassard, P., et al. 1997, ApJ, 483, 123

\bibitem[2008]{charpinet08}
Charpinet, S., van Grootel, V., Reese, D., et al. 2008, A\&A, 489, 377

\bibitem[2009]{charpinet09}
Charpinet, S., Fontaine, G., \& Brassard, P. 2009, Nature, 461, 501

\bibitem[2011a]{charpinet11a}
Charpinet, S., Fontaine, G., Brassard, P., et al. 2011a, Nature, 480, 496

\bibitem[2011b]{charpinet11b}
Charpinet, S., van Grootel, V., Fontaine, G., et al. 2011b, A\&A, 530, 3

\bibitem[2011]{copperwheat11}
Copperwheat, C. M., Morales-Rueda, L., Marsh, T. R., Maxted, P. F. L., \& Heber, U. 2011, MNRAS, 415, 1381

\bibitem[1996]{demarchi96}
de Marchi. G., \& Paresce, F. 1996, ApJ, 467, 658

\bibitem[1996]{dcruz96}
D'Cruz, N. L., Dorman, B., Rood, R. T., \& O'Connell, R. W. 1996, ApJ, 466, 359

\bibitem[2012]{deca12}
Deca, J., Marsh, T. R., \O stensen, R. H., et al. 2012, MNRAS, 421, 2798

\bibitem[1993]{dorman93}
Dorman, B., Rood, R. T., \& O'Connell, R. W. 1993, ApJ, 419, 596

\bibitem[2005]{edelmann05}
Edelmann, H., Heber, U., Altmann, M., Karl, C., \& Lisker, T. 2005, A\&A 442, 1023

\bibitem[2001]{edelmann01}
Edelmann, H., Heber, U., \& Napiwotzki, R. 2001, AN, 322, 401

\bibitem[2006]{eisenstein06}
Eisenstein, D. J., Liebert, J., Harris, H. C., et al. 2006, ApJS, 167, 40

\bibitem[2005]{fabbian05}
Fabbian, D., Recio-Blanco, A., Gratton, R. G., \& Piotto, G. 2005, A\&A, 434, 235

\bibitem[2012]{fontaine12}
Fontaine, G., Brassard, P., Charpinet, S., et al. 2012, A\&A, 539, 12

\bibitem[2010]{for10}
For, B.-Q., \& Sneden 2010, AJ, 140, 1694

\bibitem[2007]{geier07}
Geier, S., Nesslinger, S., Heber, U., et al. 2007, A\&A, 464, 299

\bibitem[2010a]{geier10a}
Geier, S., Heber, U., Edelmann, H., Morales-Rueda, L., \& Napiwotzki, R. 2010a, Ap\&SS, 329, 127

\bibitem[2010b]{geier10b}
Geier, S., Heber, U., Podsiadlowski, Ph., et al. 2010b, A\&A, 519, 25

\bibitem[2011a]{geier11a}
Geier, S., Classen, L., \& Heber, U. 2011a, ApJ, 733, L13

\bibitem[2011c]{geier11b}
Geier, S., Hirsch, H., Tillich, A., et al. 2011b, A\&A, 530, 28

\bibitem[2011d]{geier11c}
Geier, S., Schaffenroth, V., Drechsel, H., et al. 2011c, ApJ, 731, L22

\bibitem[2012]{geier12}
Geier, S., Heber, U., Edelmann, H., et al. 2012, ASP Conf. Ser., 452, 57

\bibitem[2006]{gourgouliatos06}
Gourgouliatos, K. N., \& Jeffery, C. S. 2006, MNRAS, 371, 1381

\bibitem[2003]{green03}
Green, E. M., Fontaine, G., Reed, M. D., et al. 2003, ApJ, 583, L31

\bibitem[2006]{green06}
Green, E. M., Fontaine, G., Hyde, E. A., Charpinet, S., \& Chayer, P. 2006, Baltic Astronomy, 15, 167

\bibitem[2008]{green08}
Green, E. M., Fontaine, G., Hyde, E. A., For, B.-Q., \& Chayer, P. 2008, ASP Conf. Ser.,
 392, 75

\bibitem[1999]{grundahl99}
Grundahl, F., Catelan, M., Landsman, W. B., et al. 1999, ApJ, 524, 242

\bibitem[2008]{han08}
Han, Z. 2008, A\&A, 484, 31

\bibitem[2007]{han07}
Han, Z., Podsiadlowski, P., \& Lynas-Gray, A. E. 2007, MNRAS, 380, 1098

\bibitem[2002]{han02}
Han, Z., Podsiadlowski, P., Maxted, P. F. L., Marsh, T. R., \& Ivanova, N. 
2002, MNRAS, 336, 449

\bibitem[2003]{han03}
Han, Z., Podsiadlowski, P., Maxted, P. F. L., \& Marsh, T. R. 2003, MNRAS, 341, 669

\bibitem[1986]{heber86}
Heber, U. 1986, A\&A, 155, 33

\bibitem[1999]{heber99}
Heber, U., Reid, I. N., \& Werner, K. 1999, A\&A, 348, L25

\bibitem[2000]{heber00}
Heber, U., Reid, I. N., \& Werner, K. 2000, A\&A, 363, 198

\bibitem[2009]{heber09}
Heber, U. 2009, ARA\&A, 47, 211

\bibitem[2009]{hirsch09}
Hirsch, H., \& Heber, U. 2009, JPhCS, 172, 2015

\bibitem[1984]{iben84}
Iben, I., \& Tutukov, A. V. 1984, ApJ, 284, 719

\bibitem[2000]{jeffery00}
Jeffery, C. S., \& Pollacco, D. 2000, MNRAS, 318, 974

\bibitem[2003]{kawaler03}
Kawaler, S. D. 2003, in Proceedings IAU Symposium No. 215, ed. A. Maeder \& P. Eenens, arXiv:astro-ph/0301539

\bibitem[2005]{kawaler05}
Kawaler, S. D., \& Hostler, S. R. 2005, ApJ, 621, 432 

\bibitem[2007]{kilkenny07}
Kilkenny, D., Copley, C., Zietsman, E., \& Worters, H. 2007, MNRAS, 375, 1325

\bibitem[1997]{kilkenny97}
Kilkenny, D., Koen, C., O'Donoghue, D., \& Stobie, R. S. 1997, MNRAS, 285, 640

\bibitem[2000]{kinman00}
Kinman, T. D., Castelli, F., Cacciari, C., et al. 2000, A\&A, 364, 102

\bibitem[2010]{koen10}
Koen, C., Kilkenny, D., Pretorius, M. L., \& Frew, D. J. 2010, MNRAS, 401, 1850

\bibitem[1997]{lemke97}
Lemke, M. 1997, A\&AS, 122, 285

\bibitem[2005]{lisker05}
Lisker, T., Heber, U., Napiwotzki, R., Christlieb, N., Han, Z., et al. 2005, 
A\&A, 430, 223

\bibitem[2000]{marietta00}
Marietta, E., Burrows, A., \& Fryxell, B. 2000, ApJS, 128, 615

\bibitem[2001]{maxted01}
Maxted, P. F. L., Heber, U., Marsh, T. R., North, R. C., 2001, MNRAS, 326, 139 

\bibitem[2000]{maxted00}
Maxted, P. F. L., Marsh, T. R., \& North, R. C. 2000, MNRAS, 317, L41

\bibitem[1976]{mengel76} 
Mengel J.G., Norris J., \& Gross P.G. 1976, ApJ, 204, 488 

\bibitem[1983]{michaud83}
Michaud, G., Vauclair, G., \& Vauclair, S. 1983, ApJ, 267, 256

\bibitem[2008]{michaud08}
Michaud, G., Richer, J., \& Richard, O. 2008, ApJ, 675, 1223

\bibitem[2003]{morales03}
Morales-Rueda, L., Maxted, P. F. L., Marsh, T. R., North, R. C., \& Heber, U. 2003, MNRAS, 338, 752

\bibitem[2008]{napiwotzki08}
Napiwotzki, R. 2008, ASP Conf. Ser., 392, 139

\bibitem[2001]{napiwotzki01}
Napiwotzki, R., Christlieb, N., Drechsel, H., et al. 2001, AN, 322, 411

\bibitem[2003]{napiwotzki03}
Napiwotzki, R., Christlieb, N., Drechsel, H., et al. 2003, ESO Msngr, 112, 25

\bibitem[2004a]{napiwotzki04a}
Napiwotzki, R., Karl, C. A., Lisker, T., et al. 2004a, Ap\&SS, 291, 321

\bibitem[2004b]{napiwotzki04b}
Napiwotzki, R., Yungelson, L., Nelemans, G. et al. 2004b, ASP Conf. Ser., 318, 402

\bibitem[1999]{odonoghue99}
O'Donoghue D., Koen C., Kilkenny D., Stobie R.S., 1999, ASP Conf. Ser., 169, 149

\bibitem[2010]{oestensen10}
\O stensen, R. H., Oreiro, R., Solheim, J.-E., et al. 2010, A\&A, 513, 6

\bibitem[2012]{oestensen12}
\O stensen, R. H., \& van Winckel, H. 2012, ASP Conf. Ser., 452, 163

\bibitem[2006]{pace06}
Pace, G., Recio-Blanco, A., Piotto, G., \& Momany, Y. 2006, A\&A, 452, 493

\bibitem[1983]{peterson83b}
Peterson, R. C. 1983, ApJ, 275, 737

\bibitem[1985]{peterson85}
Peterson, R. C. 1985, ApJ, 294, L35

\bibitem[1983]{peterson83a}
Peterson, R. C., Tarbell, T. D., \& Carney, B. W. 1983, ApJ, 265, 972

\bibitem[1995]{peterson95}
Peterson, R. C., Rood, R. T., \& Crocker, D. A. 1995, ApJ, 453, 214

\bibitem[2008]{politano08}
Politano, M., Taam, R. E., van der Sluys, M., \& Willems, B. 2008, ApJ, 687, L99

\bibitem[2006]{przybilla06}
Przybilla, N., Nieva, M. F., \& Edelmann, H. 2006, Baltic Astronomy, 15, 107

\bibitem[2009]{quievy09}
Quievy, D., Charbonneau, P., Michaud, G., \& Richer J. 2009, A\&A, 500, 1163

\bibitem[2006]{randall06}
Randall, S. K., Fontaine, G., Green, E. M., et al. 2006, ApJ, 643, 1198

\bibitem[2002]{recio02}
Recio-Blanco, A., Piotto, G., Aparicio, A., \& Renzini, A. 2002, ApJ, 572, L71

\bibitem[2004]{recio04}
Recio-Blanco, A., Piotto, G., Aparicio, A., \& Renzini, A. 2004, A\&A, 452, 875

\bibitem[1994]{saffer94}
Saffer, R. A., Bergeron, P., Koester, D., Liebert, J. 1994, ApJ, 432, 351

\bibitem[2000]{sills00}
Sills, A., \& Pinsonneault, M. H. 2000, ApJ, 540, 489

\bibitem[2007]{silvotti07}
Silvotti, R., Schuh, S., Janulis, R., et al. 2007, Nature, 449, 189

\bibitem[1998]{soker98}
Soker, N. 1998, AJ, 116, 1308

\bibitem[2007]{stroeer07}
Str\"oer, A., Heber, U., Lisker, T., et al. 2007, A\&A, 462, 269

\bibitem[1997]{sweigart97}
Sweigart, A. V. 1997a, in Third Conference on Faint Blue Stars, ed. A. G. D. Philip, J. Liebert, R. Saffer, \& D. S. Hayes, (Schenectady: L. Davis Press), 3

\bibitem[2002]{sweigart02}
Sweigart, A. V. 2002, Highlights of Astronomy, 12, 292 

\bibitem[2008]{telting08}
Telting, J. H., Geier, S., \O stensen, R . H., et al. 2008, A\&A, 492, 815

\bibitem[2008]{unglaub08}
Unglaub, K. 2008, A\&A, 486, 923

\bibitem[2008]{vangrootel08}
van Grootel, V., Charpinet, S., Fontaine, G., \& Brassard, P. 2008, A\&A, 483, 875

\bibitem[2010]{vangrootel10}
van Grootel, V., Charpinet, S., Fontaine, G., Green, E. M., \& Brassard, P. 2010, A\&A, 524, 63

\bibitem[2002]{vink02}
Vink, J. S., \& Cassisi, S. 2002, A\&A, 392, 553

\bibitem[1984]{webbink84}
Webbink, R. F. 1984, ApJ, 277, 355

\end{thebibliography}
\end{document}